\documentclass[11pt]{article}
\usepackage[margin=2.5cm]{geometry}
\usepackage{amsmath,amssymb}
\usepackage{graphicx}
\usepackage{float}
\usepackage[round]{natbib}
\usepackage{booktabs}
\usepackage[hidelinks,
  pdftitle={United We Stand: An Emergent Wage Phillips Curve from Collective Bargaining in an Agent-Based Goodwin Model},
  pdfauthor={Agust\'in Caputo Bugallo, Igal Kejsefman, Rodrigo Castro}
]{hyperref}
\usepackage{authblk}
\usepackage[section]{placeins}

\usepackage{tikz}
\usetikzlibrary{arrows.meta,positioning,calc}

\title{United We Stand: An Emergent Wage Phillips Curve from Collective Bargaining in an Agent-Based Goodwin Model}
\author[1,3]{Agust\'in Caputo Bugallo}
\author[2,3]{Igal Kejsefman}
\author[1,3]{Rodrigo Castro}
\affil[1]{Departamento de Computaci\'on (DC-FCEyN) and\protect\\Instituto de Ciencias de la Computaci\'on (ICC)}
\affil[2]{Instituto de Estudios de Am\'erica Latina y el Caribe (IEALC-FSOC)}
\affil[3]{CONICET and Universidad de Buenos Aires (UBA), Argentina}
\date{}

\begin{document}
\maketitle

\begin{abstract}
We show that a wage Phillips curve can emerge from collective bargaining among agents
instead of being assumed, as it is in Goodwin's growth cycle model. In an agent-based version
of that model, workers on a grid read employment within a vision radius and turn its level and
recent change into a perceived bargaining power, a union averages these perceptions to set
the wage, and a single firm accumulates capital as in Goodwin's model. A downward-sloping
curve appears only when workers organize, and it steepens the farther they see. When workers
weigh the recent change in employment, the curve opens into a loop: the same unemployment
rate brings faster wage growth while output grows above its trend than while it grows below.
Across a Latin-hypercube bank of parameter sets, the weight of that change, the workers'
memory, sets the size of this loop while the vision radius barely matters. Clustering the trajectories with k-means under a dynamic time warping distance reveals
four regimes (wide cycles, moderate cycles, near full employment and collapse); the
bargaining parameters select among the first three, and capital accumulation drives collapse.
\end{abstract}

\section{Introduction}

Goodwin's growth cycle model \citep{goodwin1967} describes capitalist growth as a conflict
over distribution. When employment is high, workers obtain faster wage increases, the wage
share rises and profits, which finance accumulation, fall; capital then grows more slowly
than productivity, employment falls, and the cycle starts again. Formally, the wage share and
the employment rate follow a predator--prey dynamics, and what closes the system is a wage
Phillips curve, the relation that \citet{phillips1958} found between unemployment and the
growth of money wages in the United Kingdom. In its linear form the curve reads
\begin{equation}
  \frac{\dot w}{w} = a(\nu - \tau),
  \label{eq:phillips}
\end{equation}
where $\nu$ is the employment rate and $a$, $\tau$ are parameters.

The curve carries the whole distributive conflict of the model, yet it enters as an
assumption, and the extensions that followed kept it that way. \citet{desai1973} added price
inflation and expectations to the wage equation, \citet{keen1995} introduced debt-financed
investment to give the model the financial instability described by Minsky, and
\citet{grasselli2012} characterized the equilibria and stability of Keen's model. In all of
them wage growth remains a given function of aggregate employment, and none says who sets
wages or what that agent knows about the labour market. The evidence suggests that the
second omission matters: across many countries, \citet{blanchflower1994} find that pay is lower where local
unemployment is higher, so wages respond to the labour market that workers actually face
rather than to a national aggregate.

Our hypothesis is that Goodwin's Phillips curve presupposes organized workers. On this
reading, wages are not set by a clearing labour market but negotiated by a union, and the strength of the
union depends on how its members read the state of employment, which each of them sees only
around itself. If this is right, the curve should appear when workers can organize and vanish
when they cannot, and its shape should depend on how much of the labour market workers see
and on how they weigh what they saw before.

Agent-based models (ABMs) can test this because they build aggregate relations from the
behaviour of interacting agents instead of postulating them \citep[see the surveys
in][]{chen2018handbook}. Macroeconomic ABMs have already derived labour-market regularities
this way. \citet{fagiolo2004} obtain Beveridge, wage and Okun curves from decentralized
matching and bargaining between individual firms and workers. The Keynes+Schumpeter model of
\citet{dosi2010schumpeter} generates endogenous growth and business cycles from
heterogeneous firms, and \citet{dosi2017} give it a labour market with firm-level wage
setting, in which more flexible labour markets raise unemployment and its volatility. In
these models many firms bargain with individual workers, in economies far richer than
Goodwin's, where the aggregate wage relation results from several mechanisms at once. We take
the opposite route: we keep Goodwin's single firm and accumulation rules
and replace only Eq.~\eqref{eq:phillips} by collective bargaining, so that any Phillips
curve in the model can be traced to the way workers perceive employment.

In our model, workers on a spatial grid see employment only within a vision radius and turn
its level and its recent change into a perceived bargaining power; a union averages these
perceptions and negotiates the wage with the firm. A downward-sloping curve appears only when
workers can organize, and it steepens with the vision radius. When workers weigh the recent
change in employment, which is their memory of the labour market, the curve opens into a
loop, and a Latin-hypercube bank of 1000 parameter sets shows that memory, not vision, sets
the size of that loop. Clustering the trajectories of the bank under a dynamic time warping distance
separates four dynamical regimes, and a classifier that predicts the regime from the
parameters ties the bargaining parameters to the kind of cycle and the firm parameters to
collapse.

Section~\ref{sec:model} presents the model and Section~\ref{sec:design} the experiments and
the measures we read from them. Section~\ref{sec:results} reports the results and discusses what locality and memory mean in
the model, and Section~\ref{sec:conclusions} concludes with extensions.

\section{The model}
\label{sec:model}

\begin{figure}[t]
  \centering
  \definecolor{emp}{HTML}{22C76A}
  \definecolor{unemp}{HTML}{F2707A}
  \definecolor{obsfill}{HTML}{F2EE6B}
  \definecolor{obs}{HTML}{F08A00}
  \definecolor{nowframe}{HTML}{7B3FD0}
  \definecolor{curve}{HTML}{0072B2}
  \definecolor{ink}{HTML}{222222}
  \begin{tikzpicture}[font=\footnotesize, x=1cm, y=1cm,
      box/.style={draw=ink, rounded corners=2pt, fill=white, align=center, inner sep=4pt, line width=0.5pt},
      head/.style={font=\footnotesize\bfseries},
      arr/.style={-{Stealth[length=2mm]}, line width=0.6pt, draw=ink},
      stage/.style={draw=ink!60, rounded corners=2pt, fill=ink!4, align=center, font=\scriptsize,
                   text width=2.75cm, minimum height=0.95cm, inner sep=2pt}]
    \begin{scope}[shift={(-0.25,0.25)}]
      \fill[unemp!30] (0,0) rectangle (3.5,-3.5);
      \foreach \x/\y in {0/0,2/0,3/0,5/0,1/1,4/1,6/1,0/2,2/2,3/2,5/2,1/3,2/3,4/3,6/3,0/4,3/4,5/4,1/5,2/5,4/5,6/5,0/6,3/6,5/6}
        \fill[emp!35] (0.5*\x+0.03,-0.5*\y-0.03) rectangle ++(0.44,-0.44);
      \draw[obs, line width=0.8pt] (0,0) rectangle (3.5,-3.5);
    \end{scope}
    \node[obs, anchor=south west, inner sep=1pt] at (-0.25,0.25) {$t-1$};
    \fill[white] (0,0) rectangle (3.5,-3.5);
    \foreach \x in {0,...,6} \foreach \y in {0,...,6}
      \fill[emp] (0.5*\x+0.03,-0.5*\y-0.03) rectangle ++(0.44,-0.44);
    \foreach \x/\y in {1/0,4/0,0/1,3/1,5/1,2/2,6/2,0/3,4/3,1/4,5/4,6/4,3/5,0/6,2/6,4/6}
      \fill[unemp] (0.5*\x+0.03,-0.5*\y-0.03) rectangle ++(0.44,-0.44);
    \draw[nowframe, line width=0.8pt] (0,0) rectangle (3.5,-3.5);
    \node[nowframe, anchor=south east, inner sep=1pt] at (3.5,0) {$t$};
    \draw[ink, line width=1.8pt] (1.0,-1.0) rectangle (2.5,-2.5);
    \node[font=\scriptsize, text=ink, fill=white, inner sep=1pt, anchor=south] at (1.75,-0.98) {neighbourhood};
    \fill[obsfill] (1.53,-1.53) rectangle (1.97,-1.97);
    \draw[obs, line width=1.2pt] (1.52,-1.52) rectangle (1.98,-1.98);
    \node[font=\scriptsize, anchor=north] at (1.75,-3.55) {$k=1$: $V_k = 9$ cells};
    \foreach \y/\fillc/\edgec/\lab in {-4.15/obsfill/obs/observing worker, -4.5/emp/emp/employed worker, -4.85/unemp/unemp/unemployed worker} {
      \fill[\fillc] (0.3,\y-0.11) rectangle ++(0.22,0.22);
      \draw[\edgec, line width=0.6pt] (0.3,\y-0.11) rectangle ++(0.22,0.22);
      \node[font=\scriptsize, anchor=west, inner sep=1pt] at (0.6,\y) {\lab};
    }
    \node[box, anchor=north west] (worker) at (4.6,0.3) {%
      \tikz[baseline=-0.1ex]\draw[obs, fill=obsfill, line width=0.6pt, sharp corners] (0,0) rectangle (1.2ex,1.2ex);\ \textbf{observing worker $i$}\\[3pt]
      level\quad $\displaystyle \ell_i = \frac{E_i(t) - (V_k{+}1)/2}{V_k+1}$\\[4pt]
      change\quad $\displaystyle c_i = \frac{E_i(t) - E_i(t-1)}{V_k+1}$\\[4pt]
      $b_i(t) = \eta\,\ell_i + (1-\eta)\,c_i$};
    \node[box, anchor=north west] (union) at (11.1,0.3) {%
      \textbf{union}\\[3pt]
      averages $b_i$ over\\the $L_t$ employed\\[3pt]
      $w_{t+1} = w_t\,(1 + \rho\,\bar b_t)$};
    \node[box, anchor=north west] (firm) at (4.6,-2.6) {%
      \textbf{firm}\\[3pt]
      $L_t = \min\!\big(N, \lfloor K_t/(sA_t) \rfloor\big)$\\[3pt]
      $Y_t = A_t L_t$, \quad $\Pi_t = Y_t - w_t L_t$\\[3pt]
      $K_{t+1} = (1-\delta)K_t + \Pi_t$, \quad $A_{t+1} = (1+\alpha)A_t$};
    \begin{scope}[shift={(12.75,-4.4)}]
      \draw[-{Stealth[length=1.5mm]}, ink] (0,0) -- (2.7,0) node[below left, inner sep=1pt, font=\scriptsize] {$u$};
      \draw[-{Stealth[length=1.5mm]}, ink] (0,0) -- (0,1.75) node[left, inner sep=1pt, font=\scriptsize] {$\frac{\Delta W}{W}$};
      \draw[ink!40, dash pattern=on 1.5pt off 1.5pt] (0,0.75) -- (2.5,0.75);
      \draw[curve, line width=0.8pt, -{Stealth[length=1.6mm]}] (2.2,0.3) .. controls (1.5,0.7) and (0.6,1.35) .. (0.2,1.6);
      \draw[curve, line width=0.8pt, -{Stealth[length=1.6mm]}] (0.2,1.6) .. controls (0.5,1.1) and (1.4,0.4) .. (2.2,0.3);
      \node[font=\scriptsize, anchor=south] at (1.35,1.75) {emergent Phillips curve};
    \end{scope}
    \draw[arr] (2.55,-1.75) to[out=0, in=180] (worker.west);
    \node[font=\scriptsize, align=center] at (4.05,-1.75) {counts\\$E_i(t)$};
    \draw[arr] (worker.east) -- node[above, font=\scriptsize, align=center] {reports\\$b_i(t)$} (union.west |- worker.east);
    \draw[arr] ($(union.south west)+(0.45,0)$) |- node[pos=0.08, left, font=\scriptsize, align=center] {bargains\\$w_{t+1}$} (firm.east);
    \draw[arr] (firm.west) to[out=180, in=0] (3.55,-2.75);
    \node[font=\scriptsize, align=center] at (4.05,-4.3) {hires at\\random};
    \begin{scope}[shift={(0,-5.95)}]
      \node[stage] (s1) at (1.2,0) {\textbf{1} release everyone,\\hire $L_t$ at random};
      \node[stage, right=0.3cm of s1] (s2) {\textbf{2} produce $Y_t$,\\reinvest profits $\Pi_t$};
      \node[stage, right=0.3cm of s2] (s3) {\textbf{3} employed workers\\count $E_i(t)$ in $V_k$};
      \node[stage, right=0.3cm of s3] (s4) {\textbf{4} union sets $w_{t+1}$\\from the mean $b_i$};
      \node[stage, right=0.3cm of s4] (s5) {\textbf{5} workers remember\\$E_i(t)$ for $t+1$};
      \foreach \a/\b in {s1/s2,s2/s3,s3/s4,s4/s5} \draw[arr] (\a) -- (\b);
      \draw[arr, ink!60] (s5.south) -- ++(0,-0.25) -| node[pos=0.25, below, font=\scriptsize, text=ink] {next period} (s1.south);
    \end{scope}
  \end{tikzpicture}
  \caption{The model. \textbf{Top:} in each period the firm hires $L_t$ workers at random on the grid. Each cell represent a worker in different states: employed (green), unemployed (red). The grid of the previous period is behind. Each observing worker (yellow) counts the employed workers in its Moore neighbourhood of radius
  $k$ (black frame), compares the count with the one
  it stored in the previous period, and reports its perceived bargaining power $b_i(t)$ to
  the union, which sets the wage of the next period. The Phillips curve on the right is not
  part of the rules: it is what the aggregate wage and unemployment trace out
  (Section~\ref{sec:results}). \textbf{Bottom:} sequence of events within a period.}
  \label{fig:model}
\end{figure}
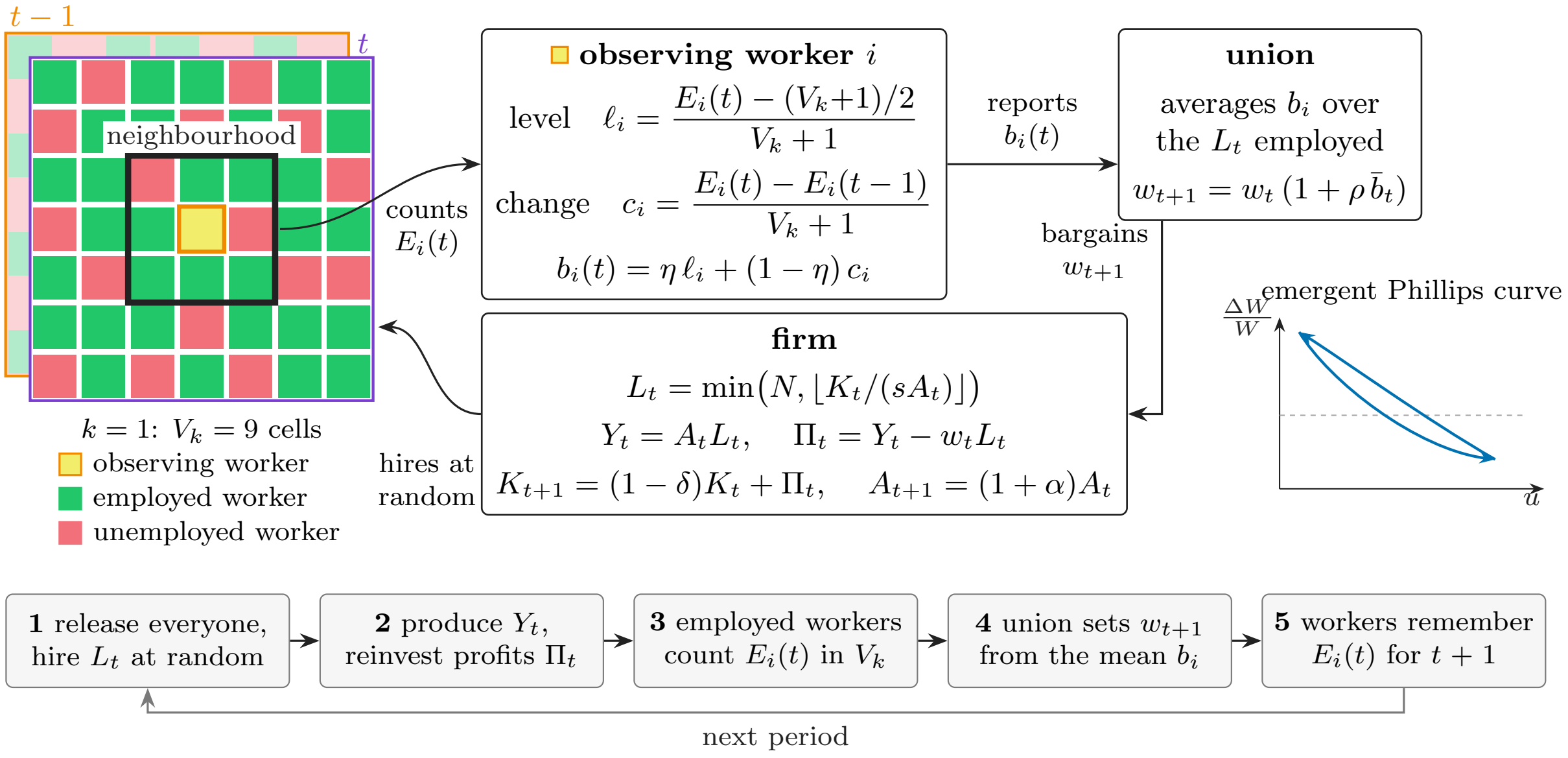

Only the wage rule separates our model from the classical Goodwin model. The economy has
$N = n \times n$ workers on a square grid with periodic boundaries, one
trade union and one representative firm that produces a single good, used both for
consumption and as capital. Time is discrete. The firm starts with capital $K_0$, labour
productivity $A_0$ and wage $w_0$. Figure~\ref{fig:model} sketches one period.

At the start of each period all workers are released, and the firm hires the number of
workers that uses its capital in full,
\begin{equation}
  L_t = \min\!\left(N, \left\lfloor \frac{K_t}{s A_t} \right\rfloor\right),
\end{equation}
where $s$ is the capital--output ratio. Workers are identical, and the firm draws them
uniformly at random from the grid. Output,
profits, capital and productivity then follow a discrete version of Goodwin's
rules \citep{goodwin1967}:
\begin{equation}
  Y_t = A_t L_t, \qquad \Pi_t = Y_t - w_t L_t, \qquad
  K_{t+1} = (1-\delta) K_t + \Pi_t, \qquad A_{t+1} = (1+\alpha) A_t,
\end{equation}
where $\delta$ is the depreciation rate and $\alpha$ the growth rate of productivity. All
profits are reinvested.

Once hired, each employed worker $i$ counts the employed workers $E_i(t)$, itself included, in the
$(2k+1)\times(2k+1)$ Moore block centred on its own cell, of size $V_k = (2k+1)^2$. Its
perceived bargaining power combines the level of local employment and its change since
the previous period,
\begin{equation}
  b_i(t) = \eta\, \frac{E_i(t) - (V_k+1)/2}{V_k+1}
         + (1-\eta)\, \frac{E_i(t) - E_i(t-1)}{V_k+1}.
  \label{eq:bargaining}
\end{equation}
The first term is positive when more than half of the worker's $V_k - 1$ neighbours are
employed; the second is its memory of the last period, the fear of losing the job after
employment fell nearby. Every worker, employed or not, stores the count around its cell at
the end of the period, so $E_i(t-1)$ is defined also for a worker who was unemployed in $t-1$. The weight $\eta \in [0,1]$ sets how much the current level counts
against that memory. We report the radius as $r = k/(n/2)$, so that $r = 1$ would let a
worker see the whole grid; in the limiting case $r = 0$ ($k = 0$) workers see no
neighbours and cannot organize, and we set their bargaining power to zero; Eq.~\eqref{eq:bargaining}
would otherwise make a worker react to its own change of status.

The union collects the bargaining power of the $L_t$ employed workers and sets the wage of
the next period,
\begin{equation}
  w_{t+1} = w_t \left(1 + \frac{\rho}{L_t} \sum_{i\ \mathrm{employed}} b_i(t)\right),
  \label{eq:wage}
\end{equation}
where $\rho$ sets how strongly the aggregate perception moves the wage. When $L_t = 0$, as at
the end of a collapse, the wage stays at $w_t$ and the wage share $\omega_t$ is not defined.

From these rules we follow four aggregate series: unemployment, wage growth, the wage share
and the growth rate of output,
\begin{equation}
  u_t = 1 - \frac{L_t}{N}, \qquad
  \frac{\Delta W_t}{W_t} = \frac{w_{t+1} - w_t}{w_t}, \qquad
  \omega_t = \frac{w_t L_t}{Y_t}, \qquad
  g_t = \frac{Y_t - Y_{t-1}}{Y_{t-1}}.
  \label{eq:observables}
\end{equation}
Wage growth is the raise negotiated at the end of period $t$, so that by
Eq.~\eqref{eq:wage} it answers to $u_t$, and $\omega_t$ uses the wage paid in $t$, so that
$w_t L_t + \Pi_t = Y_t$; we write $x_t = \Delta W_t/W_t$ in formulas. Hiring is the only
random step, and each run of the model with a given seed is a replica; the model is written
in Mesa 3.3.0 \citep{terhoeven2025mesa}.

\section{Experimental design}
\label{sec:design}

Two reference scenarios show the mechanism at work, and a parameter bank shows
how general their behaviour is. All simulations use $n = 30$ ($N = 900$ workers);
Table~\ref{tab:params} in Appendix~\ref{app:params} lists the parameters.

Both reference scenarios fix $\alpha = 0.05$, $\delta = 0.06$,
$s = 6.5$, $K_0 = 10^4$, $A_0 = 2.4$ and $w_0 = 1.4$, and each stresses one term of
Eq.~\eqref{eq:bargaining}. The first weighs the current level of employment ($\rho = 0.5$,
$\eta = 0.8$) and varies the radius over $r \in \{0, 0.07, 0.40\}$ ($k = 0, 1, 6$); the
second weighs its change ($\rho = 0.8$, $\eta = 0.4$, $r = 0.40$). Each runs 20 replicas
of 400 periods. The bank, in turn, draws 1000 parameter sets $\theta$ by Latin hypercube
sampling \citep{mckay1979lhs} over the ranges of Table~\ref{tab:params}, with $r$ rounded to
the nearest integer $k$, and runs 10 replicas of 500 periods ($t = 1, \dots, 500$) per set.

To read the business cycle out of these runs, we delimit cycles by the peaks of the wage
share: after the first 100 periods, let
$T_1 < \dots < T_{J+1}$ be the local maxima of $\omega_t$ with a topographic
prominence of at least 0.05. Cycle $j$ spans $T_j, \dots, T_{j+1}$, and the cycle length is
the mean of $T_{j+1} - T_j$. To locate each period within its cycle, we give it a
phase from the growth of output against its trend $\alpha$ and the sign of $\Delta g_t = g_t - g_{t-1}$:
\begin{equation}
  \text{phase}_t =
  \begin{cases}
    \text{growth}     & g_t > \alpha,\ \Delta g_t \geq 0,\\
    \text{recession}  & g_t > \alpha,\ \Delta g_t < 0,\\
    \text{depression} & g_t \leq \alpha,\ \Delta g_t < 0,\\
    \text{recovery}   & g_t \leq \alpha,\ \Delta g_t \geq 0.
  \end{cases}
  \label{eq:phases}
\end{equation}
The sign of $g_t$ alone cannot detect the cycle, because output rarely shrinks. At full employment
$g_t = \alpha$ exactly, so periods with $|\Delta g_t| < 10^{-9}$ keep the previous phase.

The phases describe one loop at a time. To compare loops across the bank, we measure how far
each one departs from a line by the signed area each cycle encloses in the $(u, x)$ plane (shoelace formula) over the area
of its bounding box:
\begin{equation}
  a_j = \frac{\tfrac12 \sum_{t=T_j}^{T_{j+1}} \left(u_t\, x_{t+1} - u_{t+1}\, x_t\right)}
             {\left(\max_t u_t - \min_t u_t\right)\left(\max_t x_t - \min_t x_t\right)},
  \label{eq:looparea}
\end{equation}
where the index $T_{j+1}+1$ wraps back to $T_j$, which closes the polygon, and the extrema
run over cycle $j$. Then $a_j = 0$ when
the trajectory goes back and forth along a curve and $a_j > 0$ when it turns
counter-clockwise, whatever the size of the cycle. A replica's loop area is the mean of
$a_j$; replicas with fewer than two cycles count as non-cycling, and a set cycles when at
least half of its replicas do.

The loop area summarizes a trajectory by one feature chosen in advance and exists only for
sets that cycle. To compare whole trajectories instead, we cluster the z-scored sequences $(\tilde u_t, \tilde x_t)$, $t = 100, 105, \dots,
500$, by k-means under the dynamic time warping (DTW) distance \citep{sakoe1978dtw}, which
aligns cycles of the same shape but different phase or length. We call the clusters regimes. To
choose their number we use the silhouette score; we then represent each regime by its medoid
and map the distances between trajectories with a multidimensional scaling (MDS)
\citep{perry2017narratives}. To learn which parameters place a
set in each regime, a random forest \citep{breiman2001rf} predicts the regime from the nine
parameters, and SHAP values \citep{lundberg2017shap} split its predictions among them. Appendix~\ref{app:methods} gives the definitions.

\section{Results and discussion}
\label{sec:results}

\begin{figure}[H]
  \centering
  \includegraphics[width=\linewidth]{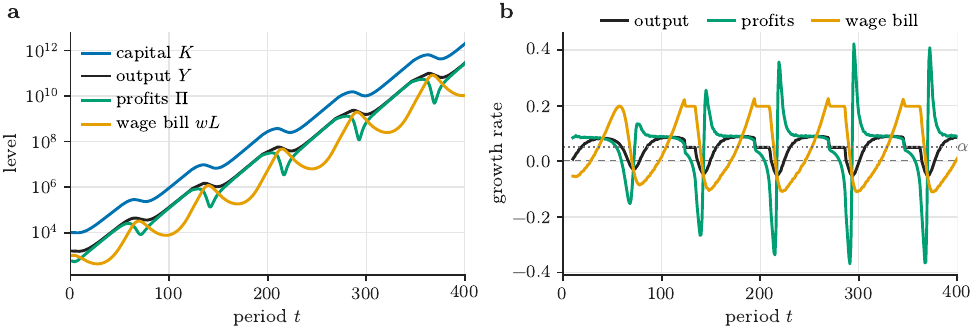}
  \caption{One run of the model, $\rho=0.5$, $\eta=0.8$, $r=0.40$ (one replica).
  (a) Capital, output, profits and wage bill, log scale. (b) Their growth rates from
  $t = 10$ (capital omitted), with the productivity growth rate $\alpha$ dotted. The plateaus at $\alpha$ are
  the periods of full employment, when output grows only through productivity.}
  \label{fig:dynamics}
\end{figure}

\begin{figure}[H]
  \centering
  \includegraphics[width=\linewidth]{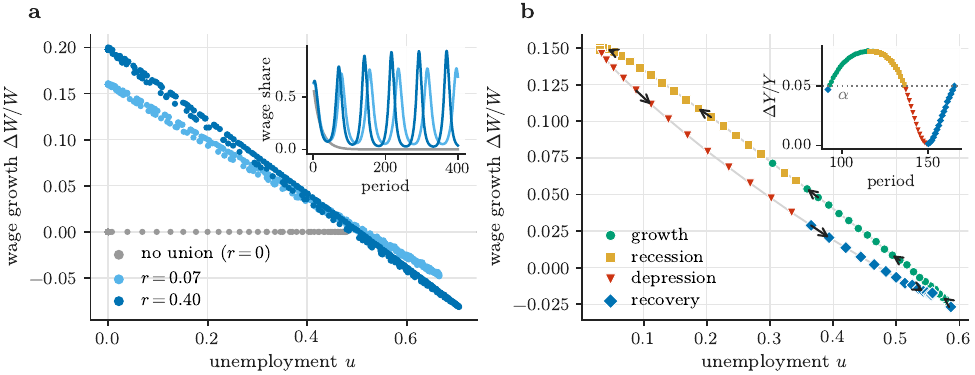}
  \caption{Wage Phillips curve with a fixed vision radius, mean of 20 replicas.
  (a) $\rho=0.5$, $\eta=0.8$ for $r=0$ (no union), $0.07$ and $0.40$; inset: wage share over
  time. (b) $\rho=0.8$, $\eta=0.4$, $r=0.40$, one cycle (periods 92--165, cut at two peaks of unemployment so that the loop starts at its right end), one marker per period by
  business-cycle phase; arrows give the direction of motion. Inset: growth rate of output,
  with its trend $\alpha$ dotted.}
  \label{fig:phillips}
\end{figure}

The model produces a growth cycle without any external shock, and its Phillips curve is the
path of that cycle in the plane of unemployment and wage growth. In one replica of the first
reference scenario with $r = 0.40$ (Fig.~\ref{fig:dynamics}), capital, output, profits and
the wage bill grow at a compound rate of about
0.05 per period after period 100, the rate of productivity, and oscillate around that trend. When unemployment is high, wages fall and profits are large, so
capital and output accelerate; the firm hires, and output growth peaks at about 0.09 per
period with unemployment near 0.4, when wages are already rising. At full employment output can only grow at $\alpha$ while the wage bill
grows at about 0.2, so the wage share climbs to about 0.9 and profits shrink. Capital then
grows more slowly than productivity, the firm needs fewer than $N$ workers, and once profits
no longer cover depreciation output contracts and unemployment reaches about 0.7. Wages
fall, profits recover and the cycle, about 76 periods long, starts again. Replicas differ
little (the standard deviation of $u_t$ across them has a median of 0.002 over time), so
Fig.~\ref{fig:phillips} shows their mean.

These cycles trace a Phillips curve only when workers can organize
(Fig.~\ref{fig:phillips}a). In the limit $r = 0$, where there is no union, wages never
move, and the wage share falls to zero as productivity grows. As soon
as workers see their neighbours a downward-sloping curve appears, with wages growing when
unemployment is below one half. A least-squares line $x_t = \beta_0 + \beta_1 u_t$ fitted
to each replica ($t \geq 2$) has slope $\beta_1 = -0.31$ for $r = 0.07$ and $-0.40$ for
$r = 0.40$, crosses zero at $u \approx 0.5$ and has $R^2 > 0.99$ in both: the farther workers see, the more strongly wages react to unemployment. With $r = 0.07$ the
cycle is a little longer, about 80 periods (inset). Nothing in the model imposes
Eq.~\eqref{eq:phillips}; the curve comes from how workers read local employment and how the
union aggregates what they read.

The curve is close to a line because, with $\eta = 0.8$, wages respond to the level of
unemployment and hardly to its history. When the change in local employment weighs more
($\eta = 0.4$, $\rho = 0.8$), the curve opens into a loop that the economy runs counter-clockwise about
every 71 periods (Fig.~\ref{fig:phillips}b). While output grows above its trend (growth and
recession in Eq.~\eqref{eq:phases}) unemployment falls along the upper branch; in
depression and recovery it rises along the lower one, where the same unemployment rate buys
a smaller raise. Workers who have just seen jobs disappear around them bargain for less.

\begin{figure}[H]
  \centering
  \includegraphics[width=\linewidth]{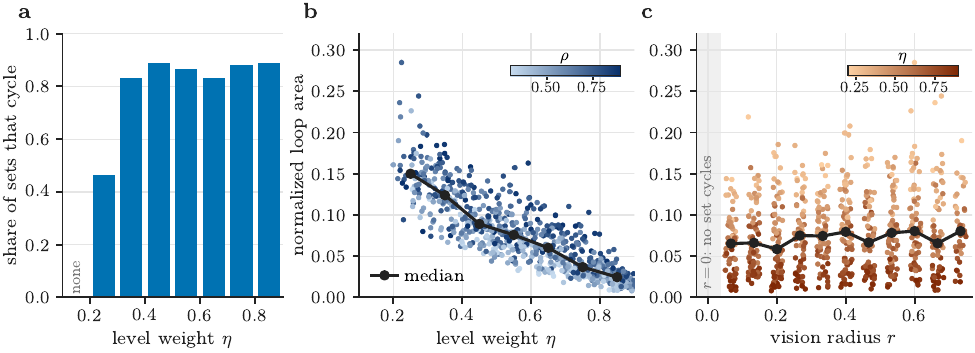}
  \caption{Path dependence across the parameter bank. (a) Share of parameter sets that
  cycle, by bin of $\eta$. (b) Normalized loop area of the $(u, \Delta W/W)$ trajectory for
  the 706 cycling sets against $\eta$, coloured by $\rho$; the black line joins the median
  per bin. (c) The same area against the vision radius $r$ (jittered within each $k$),
  coloured by $\eta$, with the median per radius.}
  \label{fig:hysteresis}
\end{figure}

Across the bank the loop has the same orientation in every set that cycles, and its size
depends on the memory of the agents (Fig.~\ref{fig:hysteresis}). Of the 1000 sets, 706 cycle. Of the other 294, 67 complete a single cycle in the window, 80
oscillate too little to count as cycles, and 147 show no swing of the wage share above 0.01,
mostly because employment settles near full employment (104) or collapses (30). No set without a union cycles, nor does any
with $\eta \leq 0.2$ (Fig.~\ref{fig:hysteresis}a); just above that threshold about half of
the sets cycle, and for $\eta > 0.3$ between 83\% and 89\%. The
normalized loop area ranges from 0.01 to 0.28 and is always positive, so every loop turns as
in Fig.~\ref{fig:phillips}b. The area shrinks as $\eta$, the weight on the level, rises: their
Spearman correlation is $-0.87$, and the median area falls from 0.15 for $\eta \in (0.2, 0.3]$ to 0.02 for
$\eta \in (0.8, 0.9]$ (Fig.~\ref{fig:hysteresis}b). A stronger wage response widens the
loop somewhat (correlation 0.29 with $\rho$), and so do a lower capital--output ratio and
lower depreciation ($-0.29$ with $s$, $-0.19$ with $\delta$), while the radius barely matters
(0.09): for every $k \geq 1$ the median area stays between about 0.06 and 0.08
(Fig.~\ref{fig:hysteresis}c).
How far workers see sets the slope of the curve; how much they remember sets how far it
opens.

\begin{figure}[H]
  \centering
  \includegraphics[width=\linewidth]{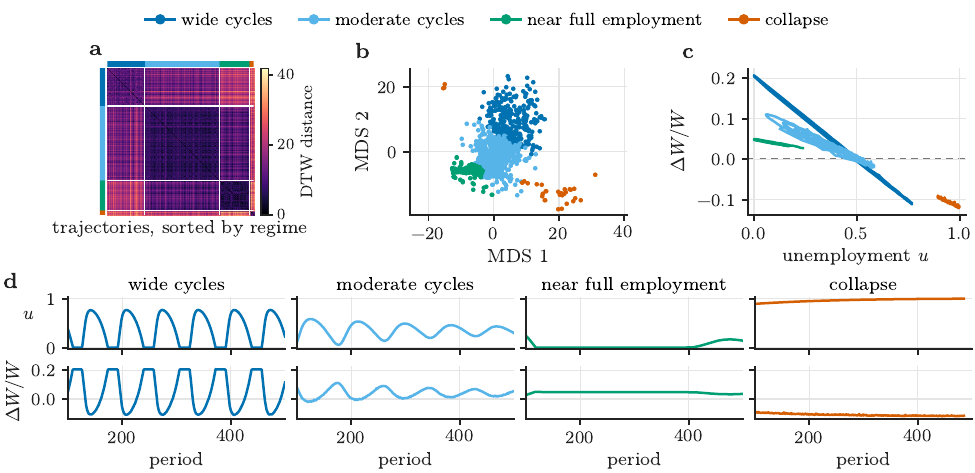}
  \caption{Dynamical regimes of the parameter bank (one replica per set). (a) DTW distances
  between trajectories, sorted by regime; the strips give the regime of each row and column.
  (b) MDS embedding of the same distances. (c) Medoid of each regime in the
  $(u, \Delta W/W)$ plane. (d) The same medoids over time, unemployment above and wage growth
  below. The collapse medoid stops when employment reaches zero.}
  \label{fig:regimes}
\end{figure}

Clustering the trajectories brings back the 294 sets that do not cycle. On one replica per set, the silhouette is highest for two clusters (0.36) and peaks again at
four (0.32); with two, collapse would join the wide cycles and the moderate cycles would be
split, so we keep four. The final model, fitted on all replicas, scores 0.34 on that sample;
the replicas of a set share a cluster in more than 99\% of the cases, and the clusters form
four blocks in the distance matrix and sit side by side in the MDS embedding
(Fig.~\ref{fig:regimes}a--b). We name them after their medoids (Fig.~\ref{fig:regimes}c--d).

The weight workers put on the level of employment orders the first three regimes, and
collapse stands apart (Table~\ref{tab:regimes}). In the 255 sets with \emph{wide cycles}, unemployment sweeps from 0 to about 0.8 along an
almost straight Phillips curve; 251 of them have $\eta > 0.5$. The 511 sets with
\emph{moderate cycles} have intermediate $\eta$ and smaller, slowly damped cycles that trace
wider loops. In the 204 sets \emph{near full employment}, unemployment stays low (its mean
over the window has a median of 0.05) and, in the sets with a union, wages grow on average
at about the rate of productivity (median absolute gap 0.002), so the wage share stays low
and capital keeps growing. This regime holds 44
of the 45 sets without a union and 100 of the 125 with $\eta \leq 0.2$, and some of its sets
may follow cycles longer than the simulated window. In the 30 sets of
\emph{collapse}, employment falls towards zero and the wage share with it.

\begin{figure}[H]
  \centering
  \includegraphics[width=\linewidth]{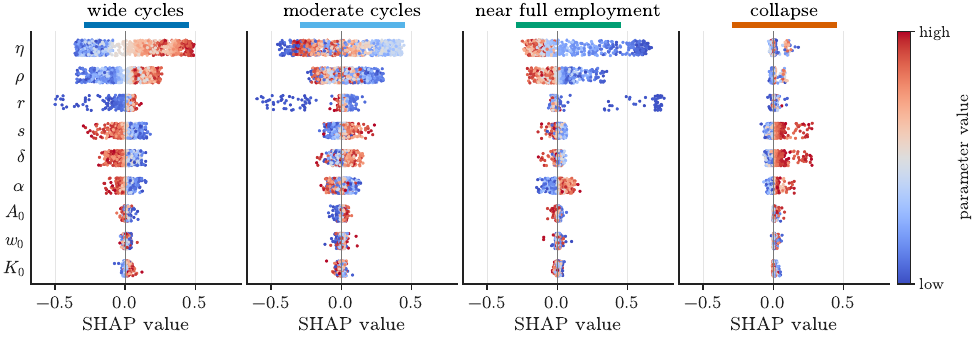}
  \caption{SHAP values of the random forest that predicts the regime, one panel per regime
  and one point per parameter set, coloured by the value of the parameter. Points to the
  right raise the predicted probability of that regime. Parameters are sorted by mean
  absolute SHAP value over all regimes.}
  \label{fig:shap}
\end{figure}

A random forest recovers the regime from the nine parameters with a cross-validated
accuracy of $0.87 \pm 0.02$, but it recovers only 4 of the 30 collapse sets. Its SHAP values (Fig.~\ref{fig:shap}) rank $\eta$ first and
$\rho$ second: high values of both push a set towards wide cycles and low values towards
near full employment, with moderate cycles in between. A short radius also pushes towards
near full employment, where the sets without a union end. Collapse depends on the firm, as
the medians of Table~\ref{tab:regimes} already show: high $\alpha$, $\delta$ and $s$, which
make it harder for capital to keep pace with productivity, drive it and also push sets out
of wide cycles. The initial conditions $K_0$,
$A_0$ and $w_0$ barely matter.

In these results, locality is the information each worker sees. Since the firm rehires its whole
workforce at random every period, no part of the grid keeps its jobs from one period to the
next, and the radius sets only how much of the labour market a worker sees; workers with a
wider radius produce a steeper Phillips curve. A variant in which the radius grows with aggregate employment,
$k_t = \mathrm{round}(k_{\min} + (k_{\max} - k_{\min})(1 - u_t))$, a simple form of union
organization that strengthens in booms, gives almost the same curve as a fixed radius for
$r \in [0.07, 0.40]$ and $r \in [0.07, 0.73]$ ($\rho = 0.5$, $\eta = 0.8$), so we kept the
radius fixed throughout.

Memory, in contrast, changes what unemployment tells about wages. The change term makes wage
growth depend on whether employment is rising or falling, and the same unemployment rate
thus maps to two rates of wage growth, one on each branch of the loop. A single line fitted to data from such an economy would mix the two
branches, and its slope would depend on which phases of the cycle the sample covers.

The regime map also separates the roles of the two sides of the model. The bargaining
parameters decide which kind of cycle the economy follows, and the large swings of Goodwin's
model need organized workers who respond to the level of employment. Whether the economy
collapses depends instead on accumulation: while the firm hires below $N$, employment holds
only if capital grows at least as fast as productivity, that is, if
$(1 - \omega_t)/s \geq \alpha + \delta$, a condition that fast productivity growth, high
depreciation and a high capital--output ratio make harder to meet.

\section{Conclusions}
\label{sec:conclusions}

A wage Phillips curve does not need to be assumed in Goodwin's model: it emerges when
workers who see part of the labour market report their bargaining power to a union, and it
disappears when they cannot organize. How far workers see sets its slope, and the weight
they give to the recent change in employment opens it into a loop, on which unemployment
alone no longer determines wage growth. That weight, followed by the strength of the wage response,
decides whether the economy cycles widely, cycles moderately or stays near full employment,
while the firm parameters drive collapse.

The model keeps one firm, one good, real wages and an inelastic labour supply, so that the
Phillips curve can be traced to the bargaining rule alone. Three extensions would relax this
design. Several firms competing for workers would let employment relations
persist, so that the employment a worker sees would depend on where firms hire and locality
would gain a geographic meaning. A heterogeneous labour force, with workers who differ in
vision, memory or bargaining weight, would test whether the aggregate curve survives when
perceptions are no longer alike. Replacing the grid by a graph, such as a network of
workplaces or social ties, would let information travel along links instead of across a
lattice, and would tie the vision radius to the structure of that network.

\bibliographystyle{apalike}
\bibliography{references}

\clearpage
\appendix
\section{Parameters and regimes}
\label{app:params}

\begin{table}[h]
  \centering
  \small
  \begin{tabular}{llcc}
    \toprule
    Symbol & Meaning & Reference & Bank range \\
    \midrule
    $\rho$ & wage response to bargaining power & 0.5 / 0.8 & $[0.3, 0.9]$ \\
    $\eta$ & weight of the level vs.\ the change of local employment & 0.8 / 0.4 & $[0.1, 0.9]$ \\
    $r$ & normalized vision radius, $k/(n/2)$ & 0, 0.07, 0.40 & $[0, 0.73]$ \\
    $\alpha$ & productivity growth rate & 0.05 & $[0.02, 0.05]$ \\
    $\delta$ & capital depreciation rate & 0.06 & $[0.03, 0.10]$ \\
    $s$ & capital--output ratio & 6.5 & $[5, 8]$ \\
    $K_0$ & initial capital & $10^4$ & $[5000, 20000]$ \\
    $A_0$ & initial productivity & 2.4 & $[1.5, 3.5]$ \\
    $w_0$ & initial wage & 1.4 & $[0.8, 2.0]$ \\
    \bottomrule
  \end{tabular}
  \caption{Model parameters, their values in the reference scenarios and their ranges in
  the Latin-hypercube bank. The grid side is $n = 30$ in all runs.}
  \label{tab:params}
\end{table}

\begin{table}[h]
  \centering
  \small
  \begin{tabular}{lrcccccccc}
    \toprule
    Regime & Sets & $\eta$ & $\rho$ & $r$ & $s$ & $\delta$ & $\alpha$ & Cycling & Loop area \\
    \midrule
    Wide cycles          & 255 & 0.74 & 0.71 & 0.42 & 6.17 & 0.059 & 0.033 & 1.00 & 0.049 \\
    Moderate cycles      & 511 & 0.46 & 0.57 & 0.37 & 6.62 & 0.068 & 0.034 & 0.83 & 0.082 \\
    Near full employment & 204 & 0.20 & 0.51 & 0.28 & 6.43 & 0.063 & 0.038 & 0.15 & 0.130 \\
    Collapse             &  30 & 0.52 & 0.64 & 0.30 & 7.72 & 0.094 & 0.043 & 0.00 & --    \\
    \bottomrule
  \end{tabular}
  \caption{The four regimes: number of parameter sets, median of the parameters that matter
  most, share of sets that cycle and median normalized loop area of the cycling ones.}
  \label{tab:regimes}
\end{table}

\section{Definitions for the regime map}
\label{app:methods}

Each replica becomes a sequence $X = (X_1, \dots, X_m)$ of points
$X_i = (\tilde u_{t_i}, \tilde x_{t_i})$ at the periods $t_i = 100, 105, \dots, 500$. The
variables are z-scored, $\tilde u = (u - \bar u)/\sigma_u$ and $\tilde x = (x - \bar x)/\sigma_x$, with
one mean and one standard deviation per variable over the whole bank, so that amplitude still
separates trajectories. Two sequences
are compared with DTW,
\begin{equation}
  D(X, Y) = \min_{\pi} \Big( \sum_{(i,l) \in \pi} \lVert X_i - Y_l \rVert^2 \Big)^{1/2},
  \label{eq:dtw}
\end{equation}
where $\pi$ runs over the warping paths: sequences of index pairs that go from $(1,1)$ to
$(m,m)$ in steps of $(1,0)$, $(0,1)$ or $(1,1)$. We use k-means under $D$, with DTW barycenters as cluster
centres \citep{petitjean2011dba}, as implemented in tslearn \citep{tavenard2020tslearn}, and
compare numbers of clusters with the mean silhouette over one replica per set, which
contrasts the mean distance $D$ from each trajectory to the rest of its cluster with the
smallest mean distance to another cluster. The final model is fitted on all replicas, and
each set takes the cluster of the majority of its replicas. The medoid of a cluster is the
member with the smallest total distance $D$ to the others, and metric multidimensional scaling
places one point per trajectory in the plane so that planar distances match $D$ in the
least-squares sense.

The random forest has 500 trees, and its accuracy is 5-fold
cross-validated. For each regime $c$, its predicted probability is split into additive SHAP
values, $f_c(\theta) = \phi_{c,0} + \sum_{p=1}^{9} \phi_{c,p}(\theta)$,
where $\phi_{c,0}$ is the mean prediction and $\phi_{c,p}$ is the Shapley value of
parameter $p$, computed exactly from the trees.

\end{document}